\documentclass{aa}
\usepackage{graphics}
\begin{document}
\title{Spectroscopic observations of the $\delta$ Scorpii binary
during its recent periastron passage\thanks{This paper is partly based on
observations obtained at the European Southern Observatory and the
South African Astronomical Observatory}
}

\author{A.~S.~Miroshnichenko\inst{1,2}
\and J.~Fabregat\inst{3}
\and K.~S.~Bjorkman\inst{1}
\and D.~C.~Knauth\inst{1}
\and N.~D.~Morrison\inst{1}
\and A.~E.~Tarasov\inst{4}
\and P.~Reig\inst{5,6}
\and I.~Negueruela\inst{7}
\and P.~Blay\inst{3}
}

\offprints{A.~S.~Miroshnichenko, \email{anatoly@physics.utoledo.edu}}

\institute{
Ritter Observatory, Dept. of Physics \& Astronomy, University of Toledo,
  Toledo, OH 43606, USA
\and Central Astronomical Observatory of the Russian Academy of Sciences
  at Pulkovo, 196140, Saint-Petersburg, Russia
\and Universidad de Valencia, Departamento de Astronom\'\i a, 46100 Burjassot,
  Valencia, Spain
\and Crimean Astrophysical Observatory and Isaac Newton Institute of
  Chile, Crimean Branch, Nauchny, Crimea, 98409, Ukraine
\and Foundation for Research and Technology-Hellas, 711 10 Heraklion, Crete,
  Greece
\and Physics Department, University of Crete, 710 33 Heraklion, Crete, Greece
\and Observatoire de Strasbourg, 11 rue de l${^\prime}$Universit\'e,
     67000 Strasbourg, France
}

\date{Received date, accepted date}

\titlerunning{Spectroscopy of $\delta$ Sco near periastron in 2000}
\authorrunning{A.~S.~Miroshnichenko et al.}

\abstract{
The bright star $\delta$ Sco has been considered a typical B0-type object
for many years. Spectra of the star published prior to 1990 showed no
evidence of emission, but only of short-term line profile variations
attributed to nonradial pulsations.
Speckle interferometric observations show that $\delta$ Sco is a binary
system with a highly-eccentric orbit and a period of $\sim$ 10.6 years.
Weak emission in the H$\alpha$ line was detected in its spectrum for the
first time during a periastron passage in 1990.
Shortly before the next periastron passage in the summer of 2000, the
binary entered a strong H$\alpha$ emission and enhanced mass loss phase.
We monitored the spectroscopic development of the Be outburst from
July 2000 through March 2001.
In this paper we present results from our spectroscopy, refine
elements of the binary orbit, and discuss possible mechanisms for the mass
loss.
}
\maketitle

\titlerunning{Spectroscopy of $\delta$ Sco in periastron}
\authorrunning{A.S.Miroshnichenko et al.}

\section { Introduction }
\label{intro}

The bright ($V$=2.3 mag.) B0.3{\sc iv} star \object{$\delta$ Scorpii}
(HD\,143275, HR\,5953) has been considered a typical B0-type object for
many years.
It was suspected of binarity nearly a century ago by Innes (\cite{in1901}),
who claimed visual detection of the secondary during a lunar occultation.
van Hoof, Bertiau, \& Deurinck (\cite{hbd63}) found its radial velocity (RV) to
vary with an amplitude of $\sim$ 15 km\,s$^{-1}$ on a time scale of 20 days.
Their data were obtained in February 1955 and showed a smooth, nearly
sinusoidal, variation over the period of the observations.
Based on 7 observations obtained in 5 different nights in 1974 and 1976,
Levato et al. (\cite{lev87}) also detected RV variations with even a larger
amplitude of $\sim$ 25 km\,s$^{-1}$ and suggested a period of 83 days.
These facts might imply that $\delta$ Sco is a single-lined spectroscopic binary,
and such RV variations are easily detectable with modern high-resolution
spectrographs, similar to those we used for this study. Below we further discuss
this problem.

Smith (\cite{sm86}) performed high-resolution spectroscopic observations
of the Si {\sc iii} 4552--4574 \AA\ lines and found short-term
line profile variability. He attributed this phenomenon to nonradial
pulsations and treated the star as single due to the absence of mass transfer.

Long-term speckle interferometric observations of $\delta$ Sco (Bedding
\cite{b93} and Hartkopf, Mason, \& McAlister \cite{hmm96}) determined
that it is a binary system with a highly-eccentric orbit.
The orbital parameters found by these authors are significantly different
even though they used almost the same data sets (see Table \ref{t1}).
Hartkopf et al. (\cite{hmm96}) noticed that the positional residuals tend to decrease
as the eccentricity approaches 1. Their orbital solution predicts very
large radial velocity (RV) variations near periastron for both components,
with the line peak separation reaching 150 km\,s$^{-1}$.
The components' brightness ratio is estimated to be 1.5--2.0 magnitudes
based on an optical interferometric observation obtained at the
Anglo-Australian Telescope on 1991 May 31/June 1 (Bedding \cite{b93}).

\begin{table*}[htb]
\caption[]{Orbital solutions for $\delta$ Sco}
\label{t1}
\begin{center}
\begin{tabular}{lllllllc}
\hline
\noalign{\smallskip}
T$_0$ &  $a$  & $e$ & $P$  & $\omega$ & $\Omega$ & $i$     & Ref.\\
      &arcsec &     & years& degrees  & degrees  & degrees &     \\
\noalign{\smallskip}
\hline
\noalign{\smallskip}
1979.3           & 0.11             & 0.82         & 10.5  & 170   & 0    & 70  &1\\
1979.41$\pm$0.14 & 0.107$\pm$0.007& 0.92$\pm$0.02& 10.58$\pm$0.08& 24$\pm$13 &
159.3$\pm$7.6 & 48.5$\pm$6.6& 2\\
\noalign{\smallskip}
\hline
\noalign{\smallskip}
2000.693$\pm$0.008 & 0.107$^{\rm a}$ & 0.94$\pm$0.01 & 10.58$^{\rm a}$ &$-1\pm$5 & 175 & 38$\pm$5 &3\\
\noalign{\smallskip}
\hline
\end{tabular}
\end{center}
\begin{list}{}
\item T$_0$ is the periastron passage epoch, $a$ is the orbital semi-major axis,
$e$ is the orbit eccentricity, $P$ is the orbital period, $\omega$ is the periastron
longitude, $\Omega$ is the node line longitude, $i$ is the orbital inclination angle.
References: 1 - Bedding (\cite{b93}), 2 - Hartkopf et al. (\cite{hmm96}), 3 - this work.
\item $^{\rm a}$ - the parameter is taken from the Hartkopf et al. (\cite{hmm96}) solution.
\end{list}
\end{table*}

Analysing profiles of Si {\sc iii} and Si {\sc iv} ultraviolet lines,
Snow (\cite{s81}) calculated a mass loss rate of $\sim 3 \times 10^{-11}$
M$_{\odot}$\,yr$^{-1}$ from the star. This value is the lowest one among
those obtained with the same technique for the stars of his sample, which
mainly contained Be stars.

Cot\'e \& van Kerkwijk (\cite{ck93}), who were searching for unidentified Be stars,
displayed an H$\alpha$ line profile of $\delta$ Sco obtained at ESO in 1990
(close to the predicted periastron passage),
which showed a weak double-peaked emission component inside
a broad photospheric absorption. Previous high-resolution spectroscopic data
obtained by Heasley \& Wolff (\cite{hw83}), apparently in 1981/2 at CFHT,
and by Grigsby, Morrison, \& Anderson (\cite{gma92}) in 1986 at KPNO showed
no emission component in H$\alpha$.

The HIPPARCOS parallax of the star is 8.12$\pm$0.88 milliarcsec (ESA \cite{esa97}),
which corresponds to a distance $D$=123$\pm$15 pc. Other important parameters
are $v$\,sin $i$=148$\pm$8 km\,s$^{-1}$ (Brown \& Verschueren \cite{bv97}),
E$_{B-V}$=0.14 mag., and log L$_{\rm bol}$/L$_{\odot}$=4.4$\pm$0.1
(with the contribution of the secondary subtracted). The star's effective
temperature as determined by different methods resulted in slightly different
values: T$_{\rm eff}$= 29760 K (Blackwell, Petford, \& Shallis \cite{bps80},
infrared flux method) and 27500 K (Heasley, Wolff, \& Timothy \cite{hwt82},
model atmospheres). Grigsby et al. (\cite{gma92}) fitted profiles
of several hydrogen and helium lines to non-LTE, line-blanketed model atmospheres
and concluded that the fits for T$_{\rm eff}$=27000 K and 28000 K (log $g$=4.0 in
both cases) are almost equally good for $\delta$ Sco.
The infrared flux method is less accurate because it involves the
bolometric stellar flux, which is poorly known for such a hot star due to
the lack of far-UV observations (see Hummer et al. \cite{hum88}).

In June 2000, S.~Otero discovered a brightening of $\delta$ Sco
by visual comparison with nearby 1st and 2nd magnitude stars. This information
along with the results of first spectroscopic observations, showing the H$\alpha$
line in emission, was reported by Fabregat, Reig, \& Otero (\cite{f00}) in
late July. Since that time the star was monitored by visual and
photoelectric photometry as well as by spectroscopy (our team)
until it became inaccessible in November. The photometric data have been recently
published by Otero, Fraiser, \& Christopher (\cite{ofc01}).
The results of our spectroscopic observations are presented in this paper.

Our observations are described in Sect. \ref{obs}, characteristics of the detected
spectral lines in Sect. \ref{results}, refinement of the orbital solution by means
of the RV measurements in Sect. \ref{orbit}, a brief discussion of possible
mechanisms responsible for the appearance of the Be phenomenon in $\delta$ Sco
in Sect. \ref{discussion}. Detailed modelling of the observed event is beyond the
scope of this paper and will be presented elsewhere.

\section{Observations and data reduction}
\label{obs}

High-resolution spectroscopic observations of $\delta$ Sco were obtained
at the 1\,m telescope of Ritter Observatory (Toledo, OH, USA) between 2000 August 5 and
2001 February 20, at the 2.6\,m Shajn telescope of the Crimean Astrophysical
Observatory (CAO, Ukraine) between 2000 July 29 and 2000 August 21, and at 1.52\,m
telescope at the ESO (La Silla, Chile) on 2000 October 23.
Low-resolution spectroscopic observations were performed with the 1.3\,m
telescope of the Skinakas Observatory (Crete, Greece) on 2000 July 19--22
and with the 1.52\,m G.~D.~Cassini telescope at the Loiano Observatory (BOL, Italy)
2000 July 26 and 29.
We also found two archived, previously unpublished, spectra. One was obtained at
Ritter on 1994 May 18 in almost the same spectral region and with the same equipment
as in 2000. The other one was obtained at the 1.9\,m telescope of the South African
Astronomical Observatory on 1998 April 16 with the GIRAFFE fiber-fed \'echelle
spectrograph, which gives a resolving power $R \simeq$ 39000 in the range 5200--10400 \AA.

At Ritter we used a fiber-fed \'echelle spectrograph with a Wright Instruments
Ltd. CCD camera. The spectra consisted of 9 non-overlapping $\sim$ 70 \AA\ orders
in the range 5285--6597 \AA\ with $R \simeq$ 26000.
The Ritter data were reduced with IRAF{\footnote
{IRAF is distributed by the National Optical Astronomy Observatories, which are
operated by the Association of Universities for Research in Astronomy, Inc., under
contract with the National Science Foundation.}. In total 24 spectra were obtained;
the exposure time for each spectrum was 20 minutes. Most of the spectra have a
signal-to-noise ratio (S/N) $\ge$100. The following lines were detected:
He {\sc i} 5876, Si {\sc ii} 6347 \& 6371 \AA, H$\alpha$ (emission), and
Na {\sc i} D$_{1,2}$ 5889 \& 5895 \AA (absorption).

At CAO the Coude spectrograph with a GEC CCD-detector (576$\times$380 pixel array)
was used. All observations were obtained in the second order of a diffraction grating
with a reciprocal dispersion of 3 \AA\,mm$^{-1}$, which corresponds to $R \sim$ 40000.
A typical exposure time for each spectrum was 4--10 minutes, giving a S/N $\ge$200.
A region of 31 \AA\ around the H$\alpha$ (5 spectra) and He {\sc i} 6678 \AA\ line
(4 spectra) was covered in all cases. Data reduction was done with the SPE code
developed by S.~G.~Sergeev at CAO.

At ESO the Fiber-fed Extended Range Optical Spectrograph (FEROS) and a thinned
back-illuminated EEV detector with 2048$\times$4096 15--$\mu$m pixels
were used. The spectrum was obtained in the range 3700--9000 \AA\ with $R \simeq
48000$ and a S/N $\simeq$ 120 near H$\alpha$; the exposure time was 3 minutes.
The spectrum was reduced with the FEROS data reduction package under MIDAS.

At Skinakas the telescope was equipped with a 2000$\times$800 ISA SITe chip CCD,
a 1301 line mm$^{-1}$ grating, and an 80 $\mu$m width slit, giving a dispersion of
1 \AA\,pixel$^{-1}$ ($R \simeq$ 3000). The range 5500--7550 \AA\ was observed on
July 19--20, while
the range 3750--5750 \AA\ was observed on July 21--22. Ten spectra with exposure
times from 1 to 5 seconds were obtained.

At BOL the telescope was equipped with the Bologna Faint Object Spectrograph and Camera
(BFOSC) and several gratings ($R \simeq$ 1000--2000 in the range 5400--8100 \AA).
On July 29 the \#9 grism in \'echelle mode was used with the grism \#10 as
cross-disperser, which gives $R \simeq$ 6000 in the region 3692--8046 \AA.
Both the Skinakas and BOL spectra were reduced using the
{\em Starlink} supported FIGARO package (Shortridge et al. \cite{smc97}).

The log of our observations is presented in Table \ref{t2}. Due to the northern
location of all the observing sites, the star was observed at large airmasses, and
strong telluric water vapour lines appear in the spectra.
Because of the high resolution of our data, most of these lines are unblended.
They were removed by interpolation between unaffected spectral regions, which were
used for construction of the normalized line profiles.

\renewcommand{\arraystretch}{0.6}
\begin{table}[htb]
\caption[]{Summary of spectroscopic observations of $\delta$ Sco in 2000--2001}
\label{t2}
\begin{center}
\begin{tabular}{ccclc}
\hline
\noalign{\smallskip}
Date  & HJD     & Obs. & Sp.region & n \\
2000  &2451000+ &      & \AA       &   \\
\noalign{\smallskip}
\hline
\noalign{\smallskip}
07/19 & 745.264&Ski &5500--7550 & 2 \\
07/20 & 746.302&Ski &5500--7550 & 3 \\
07/21 & 747.323&Ski &3750--5750 & 2 \\
07/22 & 748.302&Ski &3750--5750 & 3 \\
07/25 & 751.313&BOL &5400--8100 & 3 \\
07/28 & 754.286&CAO &H$\alpha$  & 1 \\
07/29 & 755.320&BOL &3692--8046 & 1 \\
08/04 & 761.575&Rit &5290--6600 & 1 \\
08/07 & 764.246&CAO &H$\alpha$  & 1 \\
08/07 & 764.568&Rit &5290--6600 & 1 \\
08/09 & 766.239&CAO &He {\sc i} 6678 & 1 \\
08/09 & 766.258&CAO &H$\alpha$  & 1 \\
08/10 & 767.592&Rit &5290--6600 & 1 \\
08/12 & 769.246&CAO &He {\sc i} 6678 & 1 \\
08/12 & 769.269&CAO &H$\alpha$  & 1 \\
08/13 & 770.207&CAO &He {\sc i} 6678 & 1 \\
08/20 & 777.225&CAO &H$\alpha$  & 1 \\
08/20 & 777.233&CAO &He {\sc i} 6678 & 1 \\
08/20 & 777.546&Rit &5290--6600 & 1 \\
08/24 & 781.577&Rit &5290--6600 & 1 \\
08/30 & 787.566&Rit &5290--6600 & 2 \\
08/31 & 788.548&Rit &5290--6600 & 2 \\
09/04 & 792.538&Rit &5290--6600 & 1 \\
09/06 & 794.541&Rit &5290--6600 & 1 \\
09/12 & 800.527&Rit &5290--6600 & 1 \\
09/13 & 801.525&Rit &5290--6600 & 1 \\
09/16 & 804.532&Rit &5290--6600 & 1 \\
09/17 & 805.524&Rit &5290--6600 & 1 \\
09/18 & 806.508&Rit &5290--6600 & 1 \\
09/19 & 807.506&Rit &5290--6600 & 1 \\
09/21 & 809.506&Rit &5290--6600 & 1 \\
09/26 & 814.494&Rit &5290--6600 & 1 \\
09/27 & 815.502&Rit &5290--6600 & 1 \\
09/28 & 816.503&Rit &5290--6600 & 1 \\
10/03 & 821.497&Rit &5290--6600 & 1 \\
10/23 & 841.493&ESO &3700--9000 & 1 \\
02/10$^{\rm a}$ & 951.948&Rit &5290--6600 & 1 \\
02/20$^{\rm a}$ & 961.904&Rit &5290--6600 & 1 \\
02/27$^{\rm a}$ & 968.978&Rit &5290--6600 & 1 \\
03/09$^{\rm a}$ & 978.944&Rit &5290--6600 & 1 \\
\noalign{\smallskip}
\hline
\end{tabular}
\end{center}
\begin{list}{}
\item The spectral range in column 4 expressed either in \AA\ or by the spectral
line in its centre.
\item Multiple spectra obtained during the same date (column 5) were co-added.
\item $^{\rm a}$ the spectrum was obtained in 2001
\end{list}
\end{table}

\section{ Results }
\label{results}

The Ritter observation obtained in 1994 revealed a weak emission in the
H$\alpha$ line similar to that detected by Cot\'e \& van Kerkwijk
(\cite{ck93}) in 1990. The line profile is shown in Fig. \ref{f1}a along with
a theoretical one, calculated with the code {\sc synspec} (Hubeny, Lanz \& Jeffery
\cite{hubeny}) for the atmospheric parameters from Heasley et al. (\cite{hwt82}),
the rotational velocity from Brown \& Verschueren (\cite{bv97}), and the
solar chemical composition. The equivalent width (EW) of the emission part (after
the theoretical profile subtraction) is 0.24 \AA, while the peak separation is
350 km\,s$^{-1}$. The emission component in the SAAO 1998 spectrum is hardly
recognizable.

The presence of an emission component in the He {\sc i} 5876 \AA\ line is
dubious in 1994 and is not seen in 1998 (see Fig. \ref{f1}b). Its absorption
component is deeper than the theoretical one for the described atmospheric
parameters, as was noticed by Heasley et al. (\cite{hwt82}) for other early
B-type stars (including $\delta$ Sco).
This difference might also be due to the presence of a shell component
or to an effect of microturbulence (see Leone \& Lanzafame \cite{ll98}).
In fact, the He {\sc i} line EW in the 1994 and 1998 spectra is almost the
same as the value reported by Heasley et al. (\cite{hwt82}), 0.73$\pm$0.01
\AA\ vs. 0.69 \AA. Thus, there is no evidence for a growth in the amount of
circumstellar matter around $\delta$ Sco during a significant portion of the
last binary cycle.

\begin{figure*}[htb]
\resizebox{14cm}{!}{\includegraphics{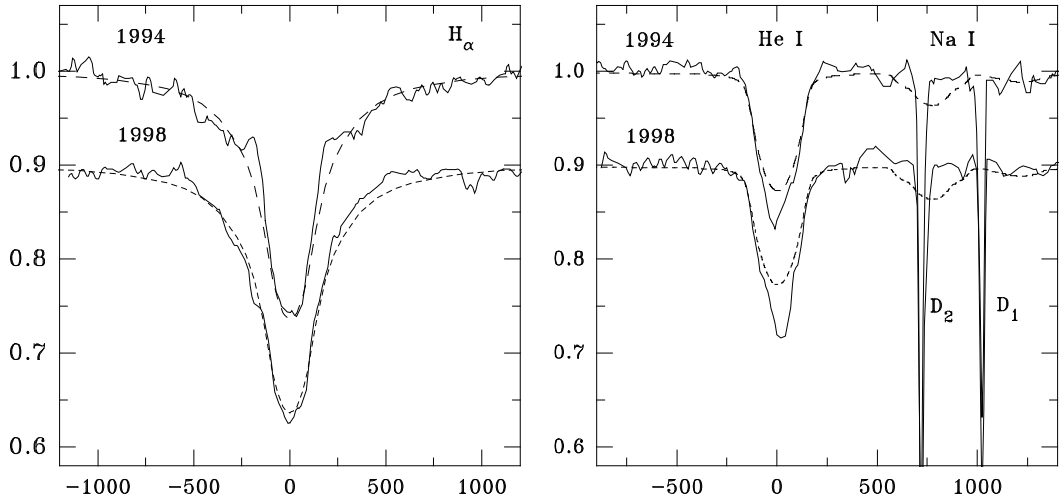}}
\caption[]{The H$\alpha$ and He {\sc i} 5876 \AA\ lines in the 1994 Ritter and
the 1998 SAAO spectra. The observational data are shown by solid
lines, while the theoretical profiles for T$_{\rm eff}$=27500 K, log $g$ = 4.0,
and v\,sin $i$ = 150 km\,s$^{-1}$ are shown by dashed lines. The intensity scale
is normalized to the level of the underlying continuum, while the heliocentric
RV are given in km\,s$^{-1}$.}
\label{f1}
\end{figure*}

All our spectra of $\delta$ Sco obtained during the 2000/1 campaign contain
emission lines.
All emission lines detected in our high-resolution spectra have double-peaked
profiles (Fig. \ref{f2}) with a mean separation of the blue and red peak of
250 km\,s$^{-1}$ (He {\sc i} and Si {\sc ii} lines) and 150 km\,s$^{-1}$
(H$\alpha$). Such a profile shape suggests that the lines are formed in a
circumstellar disk, typical for classical Be stars. The peak separation difference
can be explained by the fact that the He and Si lines are formed closer to the
star, where the disk rotates faster.

The normalized H$\alpha$ profiles obtained at CAO are systematically weaker and
narrower than those obtained at Ritter.
This might be due to the narrower spectral region observed at CAO, which contains
a small area useful for continuum determination, and a stronger contamination of the
CAO data by the telluric lines.
However, this does not affect the RVs and has no impact on our conclusions.
Parameters of the H$\alpha$ line profiles are presented in Table \ref{t3}.

Our ESO spectrum covers the whole optical range and contains information about
both photospheric and circumstellar features. The double-peaked emission is
present in all hydrogen lines up to H$_8$ and in most of the He {\sc i} lines
(see Fig. \ref{f3}).
Other species detected (e.g., O {\sc ii}, He {\sc ii}, N {\sc ii}, N {\sc iii},
Si {\sc iv}) show typical photospheric absorption profiles, which are consistent
with the mentioned fundamental parameters. No clear signs of the secondary were
found in the spectrum. This might imply that the components' temperature difference
is not large enough to produce additional detectable spectral features.
However, this suggestion needs to be verified by higher-resolution future
observations.

Our results obtained for the H$\alpha$,
He {\sc i} 5876 \AA, and Si {\sc ii} 6347 and 6371 \AA\ lines showed
that their intensities were increasing until September 12 and decreasing
after that time. The H$\alpha$ line FWHM displays a similar behaviour.
The EWs of all 4 mentioned lines showed anti-correlation with
the visual brightness of the system. Their RVs, measured by matching the
direct and reversed line profiles, were shifting to the blue
when the system was fading and vice versa (Fig. \ref{f2}ab and \ref{f4}).
At the same time, the He {\sc i} 6678 \AA\ line exhibited a small
variability of the peak strengths between August 9 and 13, but was not
detected at all on August 20 (Fig. \ref{f2}c). No sign of the Fe {\sc ii} 5317
\AA\ line was detected in the 2000 Ritter spectra. It appeared first in the ESO
spectrum and is seen in the 2001 Ritter spectra with a larger strength (Fig. \ref{f2}d).
Other Fe {\sc ii} lines, most of which have weak emission components, are also
seen in the ESO spectrum. The Na {\sc i} D lines are purely interstellar in origin
and exhibit no change in strengths or positions.

\begin{table*}[htb]
\caption[]{Characteristics of the H$\alpha$ line in the spectra of $\delta$ Sco}
\label{t3}
\begin{center}
\begin{tabular}{lclclllrcc}
\hline
\noalign{\smallskip}
Date & HJD   & Obs.  &\multicolumn{7}{c}{H$\alpha$}\\
\cline{4-10}
2000& 2451000+ &  &
 EW    & I$_{\rm b}$& I$_{\rm r}$ & I$_{\rm c}$ & RV$_{\rm mean}$ &
 $\Delta$ V$_{\rm peak}$ & FWHM\\
 & & &\AA&            &             &             & km\,s$^{-1}$    &
 km\,s$^{-1}$    &km\,s$^{-1}$\\
\noalign{\smallskip}
\hline
\noalign{\smallskip}
07/19&745.264&Ski &2.92 & 1.41 & 1.40 &      &       &     &     \\
07/20&746.302&Ski &3.15 & 1.41 & 1.41 &      &       &     &     \\
07/25&751.313&BOL &3.00 & 1.35 &      &      &       &     &     \\
07/28&754.286&CAO &2.59 & 1.46 & 1.41 & 1.30 &$-$17.0& 150 & 272 \\
07/29&755.320&BOL &3.40 & 1.42 &      &      &       &     &     \\
08/04&761.575&Rit &3.37 & 1.55 & 1.53 & 1.39 &$-$19.5& 152 & 295 \\
08/07&764.246&CAO &2.64 & 1.46 & 1.45 & 1.33 &$-$18.5& 151 & 283 \\
08/07&764.568&Rit &3.28 & 1.55 & 1.52 & 1.37 &$-$21.0& 155 & 293 \\
08/09&766.258&CAO &2.76 & 1.51 & 1.47 & 1.33 &$-$19.5& 154 & 277 \\
08/10&767.592&Rit &3.59 & 1.57 & 1.55 & 1.39 &$-$19.5& 154 & 298 \\
08/12&769.269&CAO &2.99 & 1.52 & 1.49 & 1.37 &$-$25.0& 152 & 278 \\
08/20&777.225&CAO &3.03 & 1.50 & 1.53 & 1.37 &$-$28.0& 154 & 278 \\
08/20&777.546&Rit &4.18 & 1.61 & 1.64 & 1.46 &$-$28.5& 151 & 305 \\
08/24&781.577&Rit &4.35 & 1.64 & 1.62 & 1.47 &$-$30.0& 153 & 309 \\
08/30&787.566&Rit:&4.62 & 1.65 & 1.60 & 1.50 &$-$35.0&     &     \\
08/31&788.548&Rit &4.68 & 1.64 & 1.63 & 1.54 &$-$38.5& 165 & 333 \\
09/04&792.538&Rit &5.16 & 1.73 & 1.67 & 1.57 &$-$48.5& 160 & 321 \\
09/06&794.541&Rit &5.01 & 1.72 & 1.71 & 1.57 &$-$48.0& 160 & 321 \\
09/12&800.527&Rit &5.26 & 1.77 & 1.69 & 1.57 &$-$54.0& 160 & 320 \\
09/13&801.525&Rit &4.98 & 1.71 & 1.68 & 1.56 &$-$49.0& 155 & 320 \\
09/16&804.532&Rit &4.99 & 1.69 & 1.68 & 1.57 &$-$46.0& 155 & 323 \\
09/17&805.524&Rit &4.82 & 1.67 & 1.68 & 1.56 &$-$40.5& 155 & 321 \\
09/18&806.508&Rit &4.90 & 1.68 & 1.68 & 1.55 &$-$38.5& 150 & 324 \\
09/19&807.506&Rit &4.60 & 1.71 & 1.61 & 1.55 &$-$41.5& 152 & 316 \\
09/21&809.506&Rit &4.82 & 1.73 & 1.66 & 1.60 &$-$33.5& 150 & 308 \\
09/26&814.494&Rit &4.06 & 1.65 & 1.59 & 1.49 &$-$30.5& 145 & 292 \\
09/27&815.502&Rit &3.89 & 1.63 & 1.54 & 1.49 &$-$31.0& 150 & 292 \\
09/28&816.503&Rit &3.90 & 1.64 & 1.59 & 1.48 &$-$26.0& 145 & 283 \\
10/03&821.497&Rit &3.32 & 1.62 & 1.50 & 1.44 &$-$24.5& 145 & 271 \\
10/23&841.493&ESO &2.75 & 1.48 & 1.49 & 1.37 &$-$18.0& 150 & 275 \\
02/10$^{\rm a}$&951.948&Rit &3.80 & 1.59 & 1.59 & 1.47 &$-$8.5 & 135 & 282 \\
02/20$^{\rm a}$&961.904&Rit &4.00 & 1.60 & 1.61 & 1.49 &$-$8.0 & 135 & 282 \\
02/27$^{\rm a}$&968.978&Rit &4.00 & 1.61 & 1.62 & 1.49 &$-$7.5 & 135 & 282 \\
03/10$^{\rm a}$&978.944&Rit &4.34 & 1.60 & 1.65 & 1.51 &$-$7.5 & 130 & 300 \\
\noalign{\smallskip}
\hline
\end{tabular}
\end{center}
\begin{list}{}
\item The intensities of the blue and red emission peaks normalized
to the underlying continuum are listed in columns 5 and 6, respectively;
the intensity of the central depression is given in column 7;
the RV of the overall profile is given in column 8; the peak separation
and full width at half-maximum are given in columns 9 and 10, respectively.
\item The low-resolution Skinakas data are not suitable for the RV
measurements. The BOL data do not resolve the double-peaked H$\alpha$
structure.
\item The Ritter data marked with a colon have the lowest S/N ratio.
\item $^{\rm a}$ the spectrum was obtained in 2001
\end{list}
\end{table*}

\begin{figure*}[htb]
\resizebox{14cm}{!}{\includegraphics{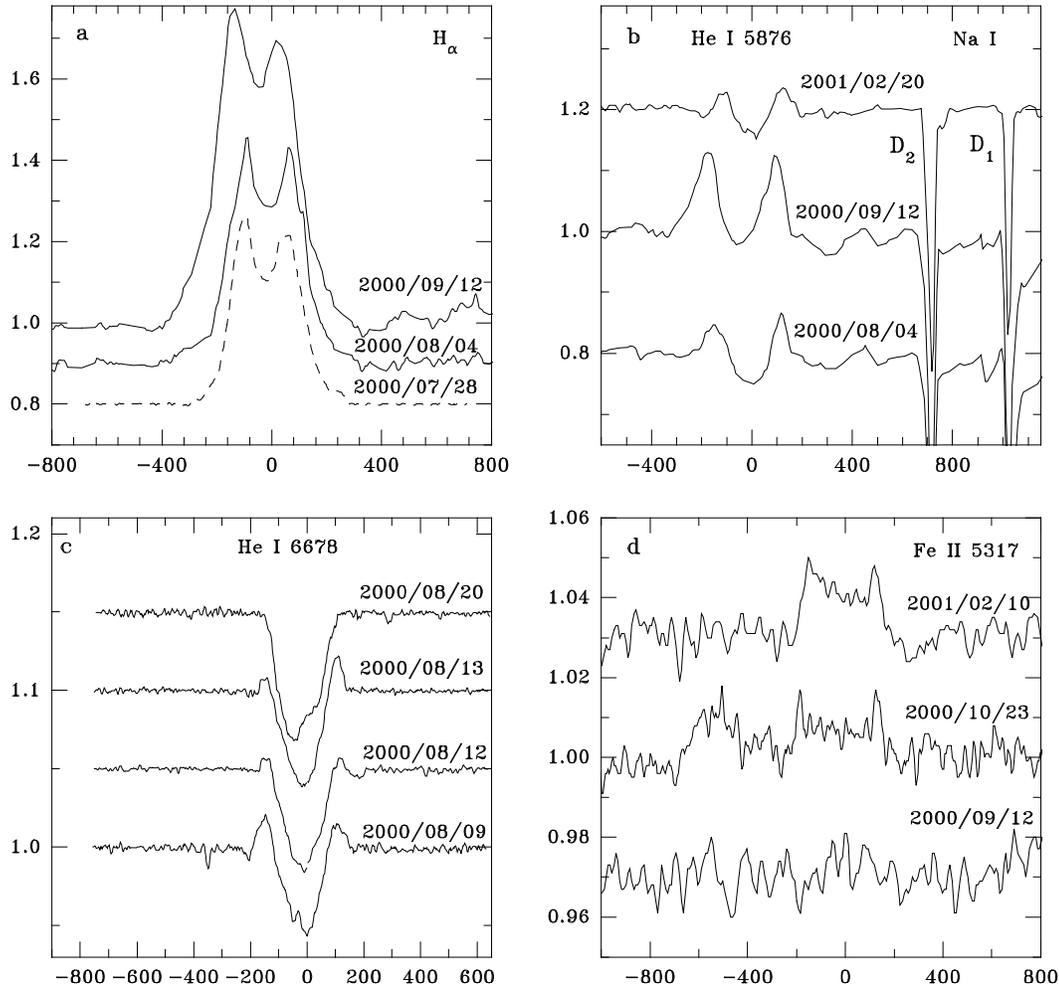}}
\caption[]{The emission line profiles of $\delta$ Sco.
a) The H$\alpha$ profiles with extreme RVs obtained at Ritter (solid lines).
The CAO H$\alpha$ profile obtained on 2000 July 28 is shown by a dash-dotted line.
b) The He {\sc i} 5876 \AA\ line profiles obtained at Ritter.
c) The He {\sc i} 6678 \AA\ line profiles obtained at CAO.
d) Time evolution of the Fe {\sc ii} 5317 \AA\ line.
The intensity and heliocentric RV are given in the same units as in Fig. \ref{f1}.
}
\label{f2}
\end{figure*}

\begin{figure*}[htb]
\resizebox{14cm}{!}{\includegraphics{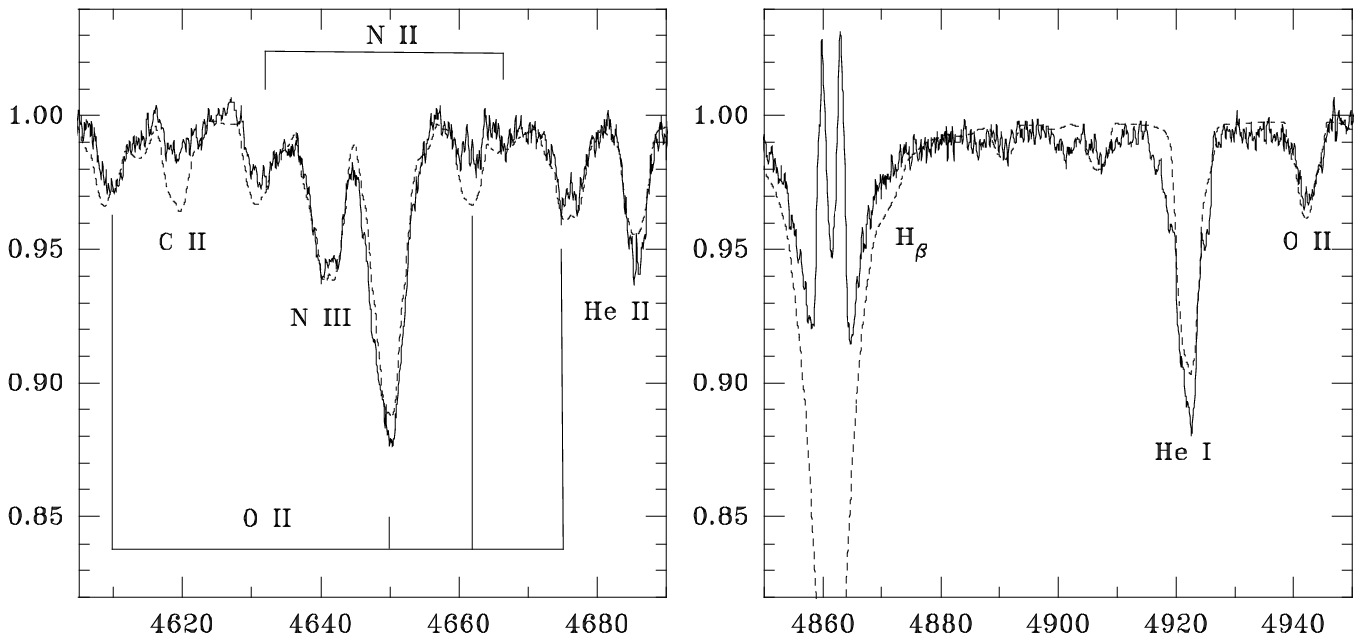}}
\caption[]{Portions of the ESO spectrum
of $\delta$ Sco obtained on 2000 October 23. The observational data are shown by a solid
line, while the theoretical profiles for T$_{\rm eff}$=27000 K, log $g$ = 4.0,
and v\,sin $i$ = 150 km\,s$^{-1}$ are shown a dashed line.
The intensity is normalized to the underlying
continuum level, while the heliocentric wavelengths are given in \AA.}
\label{f3}
\end{figure*}

\section{Refining the binary orbit}
\label{orbit}

$\delta$ Sco has been observed with the speckle interferometric technique at several
observatories for a long time (1973--1991). In total $\sim$ 30 observations were
obtained. Most of the data are collected in McAlister \& Hartkopf (\cite{mk88}),
later observations were published by Bedding (\cite{b93}), McAlister, Hartkopf,
\& Franz (\cite{mhf90}), and Fu et al. (\cite{fu97}). The data analysis has led
to two different orbital solutions presented by Bedding (\cite{b93}) and
Hartkopf, Mason, \& McAlister (\cite{hmm96}). As one can see in Fig. \ref{f5},
the data show a large scatter, suggesting that the orbital solution needs to be
tested by other observational techniques. One of the options is spectroscopy,
which gives information about the components' RVs.

We were unable to measure RVs of the photospheric lines close to periastron
(except for those in the ESO spectrum). However, the circumstellar matter
(where the emission lines originate) moves with the star and should reflect
the star's motion.
Indeed, the orbital motion of a Be star was detected using RVs of the H$\alpha$
emission line in such binaries as $\phi$ Per (Bo\v{z}i\'c et al. \cite{b95})
and $\gamma$ Cas (Harmanec et al. \cite{h2000}). Thus we think it reasonable to
use our RV measurements to constrain the orbital solution for $\delta$ Sco.
The strongest line in the spectrum, and the one that gives the most reliable RVs,
is H$\alpha$. We estimate the mean accuracy of the individual H$\alpha$ RVs to
be about 1--2 km\,s$^{-1}$.

We suggest that this line is formed in the disk around the primary component,
and its mean RV coincides with that of the star. This suggestion seems to be
fairly justified, since the line profile has almost equal peak strengths
($V/R \simeq$ 1). The latter indicates an almost axisymmetric matter distribution
in the disk.
Moreover, most of the nearly symmetric absorption lines in the ESO spectrum that
allow reliable RV measurements (such as Si {\sc iii} 4552, He {\sc i} 4143,
absoprtion wings of H$\beta$, and some others) gave values within
$\pm$2 km\,s$^{-1}$ of that from the H$\alpha$ line.

In order to compare the measured RVs with predictions from the orbital
solutions, one needs to know the systemic velocity ($v_{\rm sys}$). The RV measurements for
$\delta$ Sco published so far span a range from $-$26 to +6 km\,s$^{-1}$.
Since the orbital eccentricity is large, the RVs obtained within approximately a
year around a periastron passage are affected by the components' acceleration
and cannot be adopted as a systemic velocity. Analysis of the literature data show
that the most accurate value, which is probably not affected by the periastron
acceleration, is $-$7 km\,s$^{-1}$ (Evans \cite{e67}).
This value turned out to be in good agreement with our independent estimate from
the orbital solution fitting described below as well as with that derived from
several absorption lines in the 1998 SAAO spectrum ($-$5 km\,s$^{-1}$).

According to our data, the H$\alpha$ RV minimum took place between 2000 September
7 and 12 (see Table \ref{t3}). This is about 5 months later than the Bedding
(\cite{b93}) prediction for the periastron passage and about 1.5 months later
than the Hartkopf et al. (\cite{hmm96}) prediction. Furthermore, both previous
orbital solutions predicted a noticeably longer RV minimum (see Fig. \ref{f6}).

\begin{figure*}[htb]
\resizebox{12cm}{!}{\includegraphics{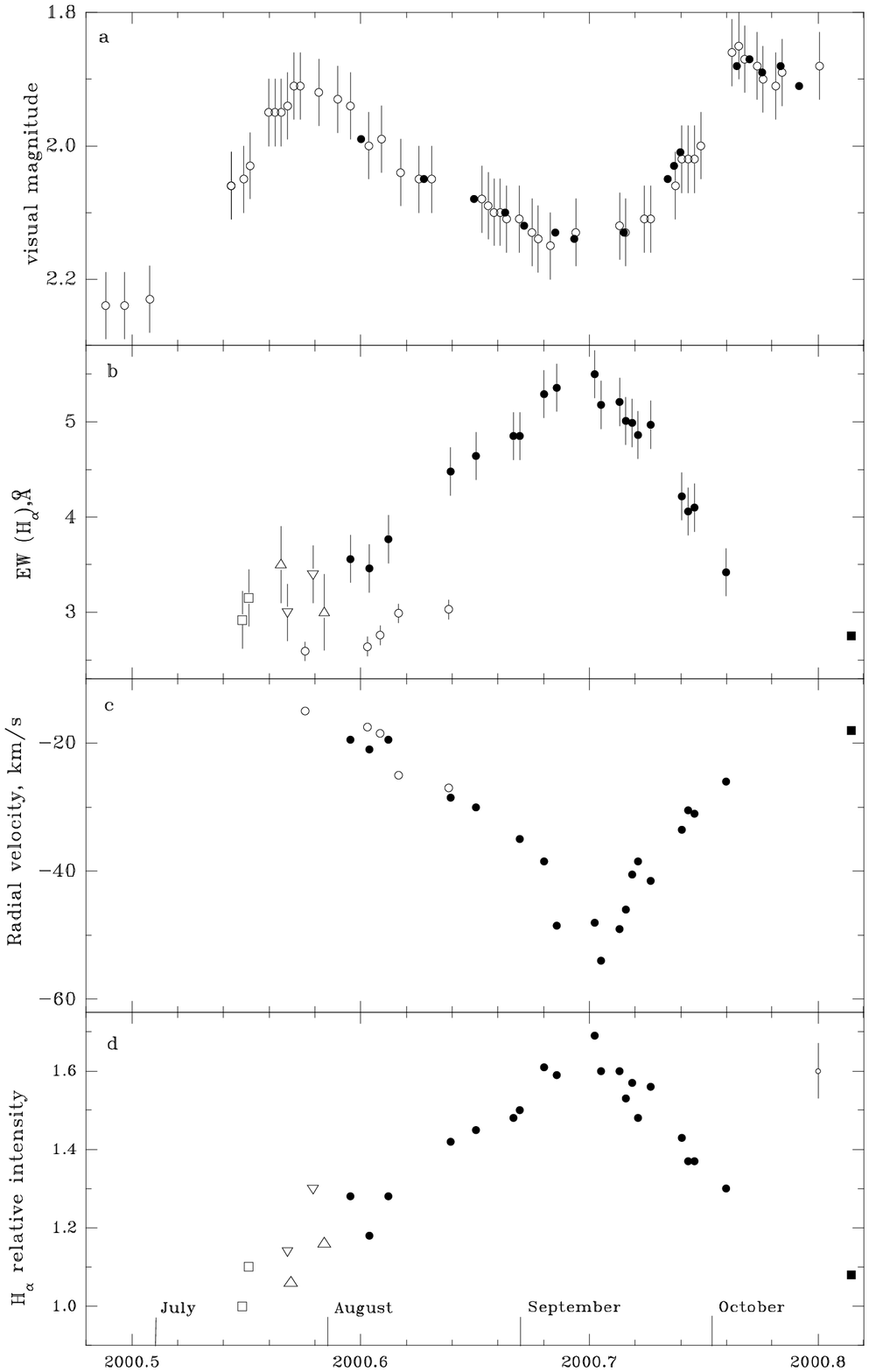}}
\caption[]{The photometric and spectroscopic variations of $\delta$ Sco
during 2000. a) The light curve from Otero et al. (\cite{ofc01}). The
visual observations are shown by open circles, and the photoelectric data by
filled circles.
The variation of the H$\alpha$ EWs, RVs (from Table \ref{t3}), and relative
flux are shown in panels b), c), and d), respectively.
The Skinakas data are shown by open squares, those obtained by French amateurs
by open upward triangles, the BOL data by open downward triangles, the CAO data
by open circles, the Ritter data by filled circles, and the ESO data by filled
squares. Uncertainties of the RVs are comparable with the symbol size,
while that of the intensities is shown in the upper right-hand corner of panel d).
}
\label{f4}
\end{figure*}

Our observations turned out to provide constraints for most of the orbital
parameters. The time coverage constrains the periastron epoch (T$_0$) within
several days. The symmetry of the RV curve suggests that the periastron longitude
($\omega$) is close to 0, while its depth depends on the components' mass ratio
($q$), the eccentricity ($e$), and the orbital inclination angle ($i$).
At the same time, $e$, $i$, and the node line longitude ($\Omega$) affect the
orbit spatial orientation, while $\Omega$ has no impact on the RV curve.
The eccentricity also controls the width of the RV curve around periastron.
The mass ratio ($q \sim$ 1.5--2.0 depending on the secondary's spectral type)
was estimated by Bedding (\cite{b93}) using the brightness ratio mentioned in
Sect. \ref{intro}. The central depression depth of the H$\alpha$ profiles
indicates that the disk inclination angle is intermediate
($\sim 30^{\circ}$--50$^{\circ}$). However, we should note here that line
profile shapes do not always give unambiguous information about the disk
inclination angle (e.g., Quirrenbach et al. \cite{q97}).
We assume that the disk plane coincides with the orbital
plane, and the primary's rotational axis is perpendicular to this plane. At least,
we do not have any evidence that the disk is tilted.

Having in hand our RV data and the speckle interferometry information, we calculated
the RV curve and the orbit projection in the plane of the sky for different sets of
the orbital parameters. The best solution is listed in the last line of Table
\ref{t1}. It is based on a study of the 6 parameters ($e$, $i$, $\omega$, T$_0$,
$q$, and $v_{\rm sys}$) space, which also provided us with the uncertainty estimates.
This study showed that $i$ and $q$ are not independent in the vicinity of the best fit.
Their relationship can be expressed as $i=(15.8\pm0.3)\,q + (10.6\pm0.5)$ and
gives $i=42^{\circ}$ for $q$=2.0 and $i=34^{\circ}$ for $q$=1.5.
The best value of $v_{\rm sys}$ is $-6\pm0.5$ km\,s$^{-1}$. It coincides with the
mean RV we measured using the 1998 SAAO spectrum taking into account a +1 km\,s$^{-1}$
correction, which follows from our orbital solution.
There are two orbital parameters we cannot estimate from our data: the
orbital period and the semi-major axis. We adopt them from the Hartkopf et al.
(\cite{hmm96}) solution, since its predicted RV curve is closer to the observed
one than that of the Bedding (\cite{b93}) solution.

Thus, our spectroscopic data improve the orbital solution for the
$\delta$ Sco system. The solution is now consistent with both the RV and
interferometric data.

\section{ Discussion }
\label{discussion}

Our new orbital solution confirms the conclusion, drawn earlier on the basis of the
interferometric data, that $\delta$ Sco is a binary system in a high-eccentricity orbit.
The presence of an additional stellar component, which may be responsible for the short-term
RV variations reported by van Hoof et al. (\cite{hbd63}) and Levato et al. (\cite{lev87}),
would result in a more complicated RV curve than the observed one.
The short-term RV amplitude is well explained by the periastron
acceleration. Perturbations in the velocity curve near periastron due to a
third body would be easily detected in the Ritter data due to their high
accuracy and dense coverage around this orbital phase.

Another important and not easily resolvable problem concerns the mass loss
and the disk formation mechanism. All the spectroscopic data available show
that the emission-line spectrum of the primary was stronger during the
current (2000) periastron passage than during the previous one. This suggests
that the amount of circumstellar matter, and hence the mass loss, grows with time.
One possibility to induce the mass loss is the mass transfer through the inner
Lagrangian point $L_1$, if at least one of the components fills its Roche lobe.
According to our orbital solution, the smallest distance between the stars at
periastron is $d$=0.06\,$a$, where $a$ is the orbital semi-major axis.
Using the components' mass ratio and the rotational parameter of the system
F$_1$=0.12 (the ratio of the primary's angular rotation velocity and the mean
orbital angular velocity), one can calculate that the equilibrium point at
periastron is located at $x_{L_1}$=0.61\,$d$ from the primary's center (Kallrath
\& Milone \cite{km98}). This point has the same meaning as $L_1$ for close
binaries in circular orbits. Since $d \sim$ 24\,R$_1$, where R$_1$ is the
primary's radius calculated from the bolometric luminosity and the spectroscopic
T$_{\rm eff}$ presented in Sect. \ref{intro}, and $x_{L_1}$=14.6\,R$_1$, the
primary is unlikely to fill its Roche lobe during a relatively fast periastron
passage. The situation is almost the same for the secondary, which is supposed to
be even smaller than the primary.

The presence of a cool secondary, filling its Roche lobe, can be ruled
out by the following arguments. As we will show below, the line
emission appeared before the optical brightening, while one would expect the
opposite in the case of the Roche lobe filling by the secondary.
Furthermore, it is easy to show that even at periastron, when an upper
limit for the secondary's radius is $\sim$10 R$_1$, the system would
brighten by $\ge$ 0.7 mag. in the $V$-band if the secondary's T$_{\rm eff}
\ge 5000 K$. If the secondary is cooler, then the overall brightness
increase would be smaller, but the absorption-line spectrum of the secondary
would be easily detectable. None of these phenomena was observed.
Thus, direct mass transfer is not expected in the present binary orbit.

As mentioned above, Smith (\cite{sm86}) found $\delta$ Sco to show nonradial
pulsations, which may result in mass loss. However, the UV spectra (Snow \cite{s81})
indicate that the mass loss from the star is extremely small. Nevertheless, $\delta$
Sco displays a detectable emission-line spectrum starting from 1990.
The data from Cot\'e \& van Kerkwijk (\cite{ck93}) and our campaign show
that the emission was observed near periastron, and that there was no
noticeable increase of the emission between the last 2 periastron passages.
Hence, one may argue that the emission strength increase correlates with the
binary period and with the periastron passage time in particular.
Since the components are very close to each other at periastron, tidal
interaction between them may become significant enough to amplify the
primary's pulsational instability and trigger an enhanced mass loss.
This is a working hypothesis which needs to be verified by modelling.
Alternatively, an unrelated growth of the disk mass (such as is seen in
other Be stars) may have been coincidentally underway at the time of this
periastron passage.

Let us now study the combined photometric and spectroscopic behaviour of $\delta$
Sco. The photometric data obtained before 1990 reveal an almost constant visual brightness,
$V=2.32\pm0.01$ mag. The only different result ($V$=2.21 mag.) was published by Hogg
(\cite{h58}). However, comparison of his photometry ($V_{\rm H}$) with the modern
Johnson system data ($V_{\rm J}$) shows that the colour difference was not properly
taken into account in Hogg's catalogue. For example, for blue stars
$V_{\rm H} \le V_{\rm J}$, while for red stars the situation is reversed.
As a result, Hogg's measurement does not differ from other values within the
inter-system translation errors.

Between 1990 January 22 and 1992 July 24 the star was monitored by the HIPPARCOS
satellite (ESA \cite{esa97}). This period covered the previous periastron passage,
and the data show no noticeable variations. The mean brightness registered by HIPPARCOS,
{$\overline V$}=2.29$\pm$0.01 mag. (translated into the Johnson system using a formula
by Harmanec \cite{h98}), is very close to the normal brightness of the star.
The HIPPARCOS value is 3 per cent brighter which is not significant due to
statistical errors of the translation. Thus, there is no evidence for any photometric
variation of $\delta$ Sco before the year 2000.

The 2000 photometric observations began visually on June 26 (Otero et al. \cite{ofc01}).
By July 4 the star was
marginally brighter than usual (m$_{\rm vis}$=2.24 mag.). The first spectroscopic
observations of $\delta$ Sco known to us in 2000 were obtained by the French amateurs
D. and S.~Morata\footnote{data accessible at\\
http://perso.wanadoo.fr/sdmorata/pages/7DelSco.html} on June 2. They show that the
H$\alpha$ line was nearly as strong as in our first spectra.
Between July 16 and 28 the star brightened up to m$_{\rm vis}$=1.9 mag. and then
experienced a $\sim$ 2-month brightness minimum (hereafter referred to as ``the dip''),
which coincided in time with the H$\alpha$ EW and intensity maximum and the line RV minimum
(see Fig. \ref{f4}).
The dip ended in the beginning of October, and the brightness remained stable
at m$_{\rm vis}$=1.89$\pm$0.02 mag. until October 18, when the star became inaccessible.
The new observing season started in late December at a similar brightness level.
Although the results of visual observers differ up to 30 per cent, the mean
star's magnitude through March 2001 is m$_{\rm vis}$=1.95$\pm$0.15.

Analysing the described variations, we emphasize the following facts:
\begin{enumerate}
\item The photometric dip and H$\alpha$ EW and FWHM maxima occurred simultaneously
and were both centered at the minimum RV, i.e. at periastron.
\item The H$\alpha$ line integrated flux also reached its maximum at periastron.
\item The H$\alpha$ peak separation slightly increased towards periastron and
gradually decreased afterwards.
\end{enumerate}

All these phenomena can be explained in the framework of the following scenario.
The system consists of a B0-type primary, surrounded by a circumstellar gaseous disk,
and a more compact secondary.
The system is not eclipsing, since the orbit is far from an edge-on orientation.
Before periastron the disk began to form (or to grow) and kept growing with time.
The rise of the H$\alpha$ intensity during the dip suggests that the amount
of matter in the disk was increasing, which, in turn, suggests an increase of
the disk's optical depth (at least before periastron).

The disk's growing size and optical depth resulted in two major effects.
The disk contribution to the overall system brightness through free-free emission
and a partial attenuation
of the primary's surface both increased with time. An interplay between these factors was
responsible for the dip's shape. In June 2000 the disk was apparently small. It produced
the H$\alpha$ emission, while its continuum emission was roughly compensated by the
attenuation so that the overall brightening of the system was small.
This also suggests that the starting point of the observed enhanced mass loss took
place not long before the first spectroscopic observation of $\delta$ Sco in 2000.

In July the disk emission grew faster than the attenuation. At some threshold point,
the decreasing separation between the stellar components limited the disk's ability
to grow further before periastron because it reaches the primary's Roche lobe size.
This implied a density increase on the side of the disk facing the secondary
which, however, smeared out quickly because of the disk's rotation, with a
period of a few days. The disk's enhanced
density caused its larger optical depth, which resulted in the overall
brightness fading. After periastron the components' separation began to increase,
allowing the disk to grow freely again. As a result, the disk's density decreased, and
the dip ended. The disk's density variations can be seen in Fig. \ref{f7}, where
larger values of the relative peak separation correspond to larger densities (see
Hanuschik, Kozok, \& Kaiser \cite{hkk88}). It is also seen that the density dropped
noticeably 5 months after periastron (triangles in Fig. \ref{f7}).
Thus, a combination of the primary's attenuation
and the disk optical brightness, which depends on both its optical depth and its spatial
extent, is capable of a qualitative explanation of the dip occurrence.

\begin{figure}[htb]
\resizebox{8cm}{!}{\includegraphics{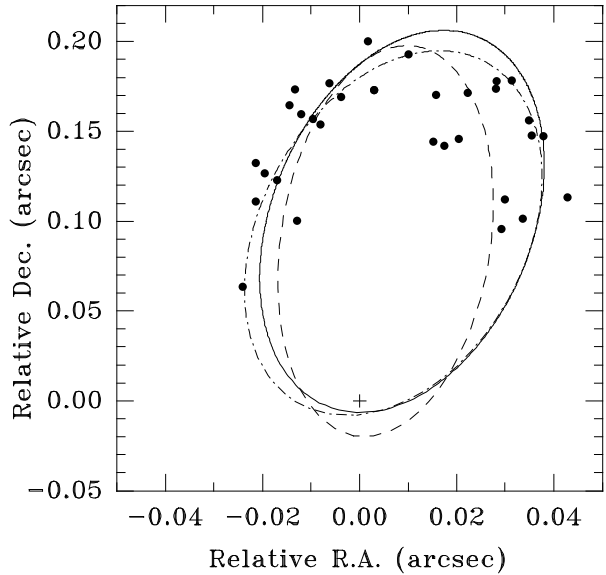}}
\caption[]{The binary orbit in the plane of the sky. The speckle interferometry
data are shown by filled circles. The primary component is placed at the origin
of the coordinate system, and is marked with a cross. The dashed line represents the
orbital solution from Bedding (\cite{b93}), the dashed-dotted line shows the solution
from Hartkopf et al. (\cite{hmm96}), and the solid line shows our solution.
}
\label{f5}
\end{figure}

\begin{figure}[htb]
\resizebox{8cm}{!}{\includegraphics{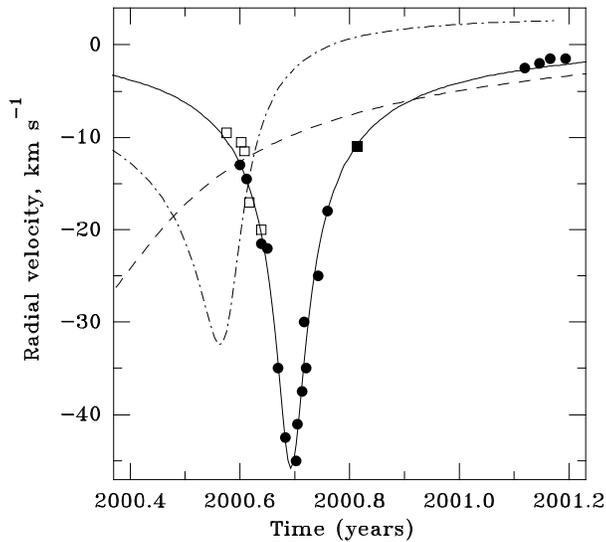}}
\caption[]{The mean RV of the H$\alpha$ profile from the Ritter (filled circles),
CAO (open squares), and ESO (filled square) data.
The measurements close in time to each other are averaged. Their uncertainties are
of the order of the point size. The observational
data are shifted by $+$6 km\,s$^{-1}$ to account for the systemic velocity.
The line types correspond to those in Fig. \ref{f5}.
}
\label{f6}
\end{figure}

\begin{figure}[htb]
\resizebox{8cm}{!}{\includegraphics{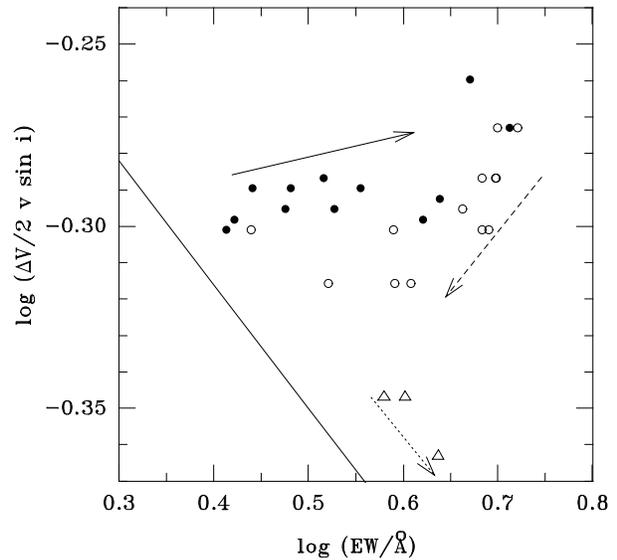}}
\caption[]{The relative peak separation of the H$\alpha$ line versus its EW.
The data obtained before periastron are shown by filled circles, those obtained after
periastron in 2000 by open circles, and the data obtained in 2001 by
triangles. The arrow-head lines show temporal trends before (solid) and
after (dashed) periastron in 2000, and in 2001 (dotted).
The solid line represents the average relationship between the displayed
parameters for Be stars found by Hanuschik (\cite{h89}).
}
\label{f7}
\end{figure}

This scenario explains most details of the system behaviour in 2000.
The observed behaviour of $\delta$ Sco during the last binary cycle
suggests that close periastron passages play an important role in the mass
loss process observed in this binary system.
However, there are a number of problems to be solved. These include
the mass loss mechanism, the nature of the secondary, and the temporal
evolution of the Fe {\sc ii} 5317 \AA\ line (see Fig. \ref{f2}d).
They will require follow up observations and modelling.

The overall behaviour of $\delta$ Sco during this period of
outburst provides a chance to see the onset of the emission line
activity, which can reveal fundamental information about the
building of stellar envelopes and the causes of the Be
phenomenon. In this sense, it is interesting to compare the
present $\delta$ Sco episode with the onset of activity in other
Be stars. One remarkable and well-studied example is \object{$\mu$ Cen},
which in the 1980's presented several activity episodes arising
from a quiescent status characterized by an absorption-line
spectrum (Peters \cite{peters}; Hanuschik et al. \cite{h93}).
A first comparison
shows that the time scales involved are very different. Outbursts
in $\mu$ Cen had a fast rise lasting 2--10 days, and a slower
decay during one to two months. Conversely, the current $\delta$
Sco outburst is still in the rising phase (by June 2001), one
year after the first detected activity. The smooth increase of
emission line strength and visual magnitude in $\delta$ Sco is
much more like the onset of activity in \object{$\gamma$ Cas} in the
1950's after several years of quiescence, leading to its current
active phase which has lasted more than 50 years by now (Cowley
\& Marlborough \cite{cm68}, Doazan et al. \cite{d83}). $\mu$ Cen also
experienced a long lived outburst which started in 1989 and led
into an active phase lasting to the present time. Superimposed on
this long term active phase, several short lived outbursts also
occurred (Rivinius et al. \cite{riv98}).
From such comparisons we might differentiate two types of
outbursts in Be stars: 1) short outbursts, produced by episodic
mass ejection, which give rise to a non-stable circumstellar
envelope that dissipates in few weeks; 2) long outbursts,
produced by continuous mass ejection, which allow the formation
of stable circumstellar envelopes that last several decades.
$\mu$ Cen presents both types of outbursts, while $\gamma$ Cas,
and so far $\delta$ Sco, present only the long lived ones.

\section{ Conclusions }

We have presented and analysed the results of our spectroscopic observations
of the $\delta$ Sco binary system during the 2000 periastron passage. The RV
curve derived from our H$\alpha$ line data was used to refine the orbital
solution previously derived from published speckle interferometric data.
In particular, we determined new values of the periastron epoch and longitude
as well as of the orbital inclination angle. The eccentricity value turned out
to be close to that in the solution of Hartkopf et al. (\cite{hmm96}).
The new orbital solution indicates that the binary is detached, and no direct
mass transfer takes place. A hypothesis that the primary's nonradial pulsations
are amplified at periastron and are responsible for the mass loss is suggested.

Our data, supplemented by the results of visual photometry, suggest
a scenario which explains most of the phenomena observed during the 2000/1
$\delta$ Sco observing campaign. For example, the photometric dip centered at
periastron and certain changes of the line profiles are attributed to an
attenuation of the primary by its disk as well as to variations of the
disk's optical depth, density, and size.

We plan to continue studying this remarkable system.
This will include follow up spectroscopic, photometric, and polarimetric
observations as well as efforts to model the observed phenomena in more
detail.

\begin{acknowledgements}
We thank W.~J.~Fischer and I.~Mihailov for their help in obtaining spectra
at Ritter.
A.~M. and K.~S.~B. acknowledge support from NASA grant NAG5-8054.
K.~S.~B. is a Cottrell Scholar of the Research Corporation,
and gratefully acknowledges their support.
Support for observational research at Ritter Observatory has been
provided by The University of Toledo, by NSF grant AST-9024802 to
B.~W.~Bopp, by an AAS small grant (NASA funding), and by a grant
from the Fund for Astrophysical Research.
Technical support is provided by R.~J.~Burmeister.
P.~R. acknowledges partial support from the European Union Training and
Mobility of Researchers Network Grant ERBFMRX/CT98/0195.
The G.~D.~Cassini telescope is operated at the Loiano Observatory by the
Osservatorio Astronomico di Bologna.
The SAAO spectrum was obtained as part of a different project, which results
have not been published yet. I.~N. would like to thank D.~A.~H.~Buckley
for his help with the observations at SAAO and J.~J.~Rodes for
pre-processing of the data.
This research has made use of the SIMBAD database operated at CDS,
Strasbourg, France.
\end{acknowledgements}

\listofobjects

\begin{thebibliography}{}

\bibitem[1993]{b93}
Bedding, T.R. 1993, AJ, 106, 768
\bibitem[1980]{bps80}
Blackwell, D.E., Petford, A.D., \& Shallis, M.J. 1980, A\&A, 82, 249
\bibitem[1995]{b95}
Bo\v{z}i\'c, H., Harmanec, P., Horn, J., Koubsk\'y, P., Scholz, G., McDavid, D.,
Hubert, A.-M., \& Hubert, H. 1995, A\&A, 304, 235
\bibitem[1997]{bv97}
Brown, A.G.A., \& Verschueren, W. 1997, A\&A, 319, 811
\bibitem[1993]{ck93}
Cot\'e, J., \& van Kerkwijk, M.H. 1993, A\&A, 274, 870
\bibitem[1968]{cm68}
Cowley, A.P., \& Marlborough, J.M. 1968, PASP, 80, 42
\bibitem[1983]{d83}
Doazan, V., Franco, M., Rusconi, L., Sedmak, G., \& Stalio, R. 1983, A\&A,
 128, 171
\bibitem[1997]{esa97}
ESA 1997, The Hipparcos and Tycho Catalogues ESA SP-1200
\bibitem[1967]{e67}
Evans, D.S. 1967, In Determination of Radial Velocities and Their Applications,
eds. A.H.Batten and J.F.Heard, Proc. of the IAU Symp. 30, Academic Press,
London, p.57
\bibitem[2000]{f00}
Fabregat, J., Reig, P., \& Otero, S. 2000, IAUC 7461
\bibitem[1997]{fu97}
Fu, H.H., Hartkopf, W.L., Mason, B.D. et al. 1997, AJ, 114, 1623
\bibitem[1992]{gma92}
Grigsby, J.A., Morrison, N.D., \& Anderson, L.S. 1992, ApJS, 78, 205
\bibitem[1988]{hum88}
Hummer, D.G., Abbott, D.C., Voels, S.A., \& Bohannan, B. 1988, ApJ, 328, 704
\bibitem[1988]{hkk88}
Hanuschik, R., Kozok, J.R., \& Kaiser, D. 1988, A\&A, 189, 147
\bibitem[1989]{h89}
Hanuschik, R. 1989, Ap\&SS, 161, 63
\bibitem[1993]{h93}
Hanuschik, R.W., Dachs, J., Baudzus, M., \& Thimm, G. 1993, A\&A 274, 356
\bibitem[1998]{h98}
Harmanec, P. 1998, A\&A, 335, 173
\bibitem[2000]{h2000}
Harmanec, P., Habuda, P., \v{S}tefl, S., et al. 2000, A\&A, 364, L85
\bibitem[1996]{hmm96}
Hartkopf, W.L., Mason, B.D., \& McAlister, H.A. 1996, AJ, 111, 370
\bibitem[1982]{hwt82}
Heasley, J.N., Wolff, S.C., \& Timothy, J.G. 1982, ApJ, 262, 663
\bibitem[1983]{hw83}
Heasley, J.N., \& Wolff, S.C. 1983, ApJ, 269, 634
\bibitem[1958]{h58}
Hogg, A.R. 1958, Mount Stromlo Obs. Mimeo. No. 2
\bibitem[1995]{hubeny}
Hubeny, I., Lanz, T., \& Jeffery, C.S. 1995, Synspec -- A User's Guide, Version 36
\bibitem[1901]{in1901}
Innes, R.T.A. 1901, MNRAS, 61, 414
\bibitem[1998]{km98}
Kallrath, J., \& Milone, E.F. 1998, Eclipsing binary stars. Modeling and Analysis.,
Springer
\bibitem[1994]{k94}
Kurucz, R.L. 1994, Smithsonian Astrophys. Obs., CD-ROM No. 19
\bibitem[1998]{ll98}
Leone, F., \& Lanzafame, A.C. 1998, A\&A, 330, 306
\bibitem[1987]{lev87}
Levato, H., Malaroda, S., Morrell, N., \& Solivella, G. 1987, ApJS, 64, 487
\bibitem[1988]{mk88}
McAlister, H.A., \& Hartkopf, W.L. 1988, CHARA contribution 2
\bibitem[1990]{mhf90}
McAlister, H.A., Hartkopf, W.L., \& Franz, O.G. 1988, AJ, 99, 965
\bibitem[2001]{ofc01}
Otero, S., Fraiser, B., \& Christopher, L. 2001, IBVS 5026
\bibitem[1986]{peters}
Peters, G.J., 1986, ApJ, 301, L61
\bibitem[1997]{q97}
Quirrenbach, A., Bjorkman, K.S., Bjorkman, J.E., et al. 1997, ApJ, 479, 477
\bibitem[1998]{riv98}
Rivinius, T., Baade, D., \v{S}tefl, S., et al. 1998, A\&A 333, 125
\bibitem[1997]{smc97}
Shortridge, K., Meyerdicks, H., Currie, M. et al. 1997, Starlink User Note
86.15, R.A.L.
\bibitem[1986]{sm86}
Smith, M.A. 1986, ApJ, 304, 728
\bibitem[1981]{s81}
Snow, T.P., Jr. 1981, ApJ, 251, 139
\bibitem[1963]{hbd63}
van Hoof, A., Bertiau, F.C., \& Deurinck, R.S.J. 1963, ApJ, 137, 824
\end{thebibliography}
\end{document}